\documentclass[AMA,Times1COL]{WileyNJDv5} %STIX1COL,STIX2COL,STIXSMALL

\articletype{Article Type}%

\received{Date Month Year}
\revised{Date Month Year}
\accepted{Date Month Year}
\journal{Journal}
\volume{00}
\copyyear{2023}
\startpage{1}

\graphicspath{ {figures/} }
\DeclareMathOperator{\pbeta}{\texttt{pbeta}}
\DeclareMathOperator{\qbeta}{\texttt{qbeta}}
\DeclareMathOperator{\fft}{\texttt{fft}}
\DeclareMathOperator{\ifft}{\texttt{ifft}}
\let\Pr\relax
\DeclareMathOperator{\Pr}{\mathbb{P}}
\newcommand{\dx}{\Delta x}
\renewcommand{\bm}[1]{\boldsymbol{#1}}

\begin{document}

% \title{Nonparametric Tolerance Limits and Sample Size With Biased Sampling}
\title{Sample Size Determination Under Selection Bias: Robust Tolerance Limits for Prevalent Cohort Data}

\author[1]{James H. McVittie}

\author[2]{Martin Lysy}

\author[3]{Masoud Asgharian}

\authormark{MCVITTIE \textsc{et al.}}
\titlemark{Nonparametric Tolerance Limits and Sample Size With Biased Sampling}

\address[1]{\orgdiv{Department of Mathematics and Statistics}, \orgname{University of Regina}, \orgaddress{\state{Saskatchewan}, \country{Canada}}}

\address[2]{\orgdiv{Department of Statistics and Actuarial Science}, \orgname{University of Waterloo}, \orgaddress{\state{Waterloo}, \country{Canada}}}

\address[3]{\orgdiv{Department of Mathematics and Statistics}, \orgname{McGill University}, \orgaddress{\state{Montreal}, \country{Canada}}}

\corres{Corresponding author: Martin Lysy, \email{mlysy@uwaterloo.ca}}

%\presentaddress{3737 Wascana Parkway, Regina, Saskatchewan, S4S 0A2}

%\fundingInfo{Text}
%\JELinfo{ejlje}

\abstract[Abstract]{Tolerance limits have received considerable attention in the statistical literature, with applications reaching far beyond their initial role in quality control. The well-known formula of Scheff\'e and Tukey (1944) establishes a simple, distribution-free relation between sample size and population coverage by two given order statistics and a given confidence level. A key requirement in applying this formula is the availability of an unbiased, representative sample from the population of interest. However, as it often happens in biological and medical applications, various logistical constraints may preclude the possibility of obtaining an unbiased sample. We derive extensions of this formula which accommodate a large class of biased sampling schemes including weight bias and censoring. The modified formulae are validated through a simulation study and compared to its unmodified counterpart. We illustrate the use of the modified formulae using the partially observed failure times for individuals with dementia using data collected from the Canadian Study of Health and Aging.}


%\jnlcitation{\cname{%
%\author{Taylor M.},
%\author{Lauritzen P},
%\author{Erath C}, and
%\author{Mittal R}}.
%\ctitle{On simplifying ‘incremental remap’-based transport schemes.} \cjournal{\it J Comput Phys.} \cvol{2021;00(00):1--18}.}

\maketitle

\renewcommand\thefootnote{}
\footnotetext{\textbf{Abbreviations:} CDF, cumulative distribution function; PDF, probability density function; CSHA, Canadian Study of Health and Aging; NPMLE, nonparametric maximum likelihood estimator; FFT, fast Fourier transform; iid, independent and identically distributed.}

\renewcommand\thefootnote{\fnsymbol{footnote}}
\setcounter{footnote}{1}

\section{Introduction}\label{sec:intro}

The determination of a sufficient sample size in the design of a medical study is crucial in making  statistically significant claims through hypothesis tests or to ensure the sample captures the majority of the population distribution. The calculation of the sample size must be completed carefully as too few observations can result in low power of the hypothesis tests or extremely wide confidence intervals. Furthermore, in most applied settings, there are logistical or cost constraints to acquiring an arbitrary number of observations for a given study. 

An early study examining the sample size problem based on tolerance limits and coverage of the population distribution can be attributed to Shewhart\cite{shewh} although Wilks\cite{wilks} is generally credited with the theoretical framework which stemmed other advances in parametric and nonparametric inference. In general terms, the sample size problem of Wilks was based on determining the sample size required for a particular proportion of the population to lie between predetermined order statistics. Scheff\'e and Tukey\cite{schef} proposed a solution to the sample size problem with a  nonparametric sample size formula which was distribution-free and did not require any assumptions on the population.     

The key assumption in the sample size formula of Scheff\'e and Tukey is having access to a random, unbiased sample from the population of interest. In practice, however, it often happens that certain logistical constraints preclude the possibility of obtaining such a sample. Some classical examples of biased sampling designs in the literature include Wicksell's corpuscle problem\cite{wicks}, Fisher's visibility bias\cite{fishe}, Neymann's incident-prevalence bias\cite{neyma} and Cox's truncation problem\cite{cox}. Manifestations of these examples are abundant in survival data collected for epidemiological studies\cite{wolfs}. Lachin and Foulkes\cite{lachin} discussed how the type of patient entry, losses in follow-up and noncompliance producing right-censored failure time data can impact a sample size calculation whereas Liu et al.\cite{liu} extended the sample size formula for a collection of left-truncated right-censored failure time data. Recently, McVittie\cite{mcvit} proposed sample size formulae for when different collections of current lifetimes, residual lifetimes, right-censored failure times and left-truncated right-censored failure times are collected under the framework of a single study. We note however that these recent approaches with right-censored and left-truncated right-censored failure time data rely on the assumption that the population failure time distribution is exponentially distributed. In this paper, we propose sample size formulae for general sampling designs producing biased/censored data under the coverage problem framework. Although we require a priori knowledge of the population distribution, we note that our proposed methods are not explicitly restricted to only the exponential distribution.  

In Section~\ref{sec:meth}, we present two different modifications of Scheff\'e and Tukey's formula which account for many patterns of biased sampling, including the ones presented above. We will also show that the modified sample size formulae cannot be distribution free as it was for the case of unbiased sampling, and thus some knowlege of both the observed and target distributions are required. Fortunately, the nature of the bias is often known such that calculations can be carried through with only a random sample from the biased population through a secondary pilot study. Specifically, we highlight how our coverage and sample size estimates can be computed nonparametrically using nonparametric estimators of the unbiased/biased population distribution functions. Through various simulation studies in Section~\ref{sec:sim}, we highlight how the original formulae of Scheff\'e and Tukey yield the incorrect sample size and coverage probabilities when applied to biased populations and examine the performance of our proposed modified sample size and coverage formulae when applied to  length-biased right-censored failure time data. We apply these procedures to survival data collected from the Canadian Study of Health and Aging (CSHA) in Section~\ref{sec:appl} with some concluding remarks given in Section~\ref{sec:disc}.

\section{Sample Size Formulas for Unbiased Sampling}\label{sec:back}

Let $X$ denote the random variable associated with the target population of interest, $F$ denote the corresponding cumulative distribution function (CDF), and let $X_1 \le \cdots \le X_n$ denote the order statistics of the random sample drawn according to $F$. Then, the sample size problem proposed by Wilks\cite{wilks} is equivalent to solving for $n$ in the inequality
\begin{equation}
\Pr\left[ F(X_{n+1-m}) - F(X_r) \geq q \right] \geq 1 - \alpha,
\label{sampsizeeq}
\end{equation}
for some predetermined proportion $q$, probability $1 - \alpha$ and order statistic indices $r$ and $m$. Through a standard distributional argument, the difference inside the probability statement~\eqref{sampsizeeq} is Beta distributed: 
\begin{equation}
F(X_{n+1-m}) - F(X_r)  
\sim \text{Beta}(n+1-k, k),
\label{betafree}
\end{equation}
where $k = r + m$ and~\eqref{betafree} is independent of the form of $F$. Based on this relation, the sample size $n$ solving~\eqref{sampsizeeq} can readily be computed using standard root-finding algorithms by solving for $n$ in the equation
\begin{equation}
\pbeta(1-q; k, n+1-k) = 1-\alpha,    
\label{sampsizeeqbeta}
\end{equation}
where $\pbeta(x; a, b)$ denotes the CDF of a $\text{Beta}(a, b)$ distribution evaluated at $x$.  Scheff\'e and Tukey\cite{schef} derive an accurate closed-form approximation to the solution of~\eqref{sampsizeeqbeta} given by
\begin{equation}
n \cong \frac{1}{4}x_{\alpha} \frac{1+q}{1-q} + \frac{1}{2}(k-1), 
\label{scheffesamp}
\end{equation}
where $x_\alpha$ is the $1 - \alpha$ quantile of a Chi-Square distribution with $2k$ degrees of freedom.

The developments above can also be used to solve the tolerance limit ``coverage problem'', namely, to solve for $q$ in equations~\eqref{sampsizeeq} and~\eqref{sampsizeeqbeta} for fixed $\alpha$, $r$, $m$, and $n$.  
% by isolating $q$ in~\eqref{scheffesamp}.  
An exact solution is obtained by solving for $q$ in~\eqref{sampsizeeqbeta}:
\begin{equation}
q = 1 - \qbeta(1-\alpha; k, n+1-k),
\label{unbiasedcov}
\end{equation}
where $\qbeta(p; a, b)$ is the inverse CDF of a $\text{Beta}(a,b)$ distribution evaluated at $p$.  Alternatively, one may solve for $q$ in~\eqref{scheffesamp} to obtain the Scheff\'e-Tukey approximation:
\begin{equation}
q \cong \frac{n - \frac{1}{2}(k-1) - \frac{1}{4} x_{\alpha}}{n - \frac{1}{2}(k-1) + \frac{1}{4}x_{\alpha}}.
    \label{scheffecov}
\end{equation}

\section{Extensions to Biased Sampling}\label{sec:meth}

An attractive feature of the Scheff\'e-Tukey formulae~\eqref{scheffesamp} and~\eqref{scheffecov} is that they are distrution-free, provided that we have access to an unbiased random sample from the target population distribution $F$.  In many epidemiological applications, however, prevalent/biased sampling designs are used instead of incident/unbiased sampling designs due to the availability of study resources in acquiring a sufficient number of failure events for conditions with long survival times. One sampling scheme, adopted through the Canadian Study of Health and Aging\cite{wolfs}, screened subjects with prevalent dementia, retrospectively recorded their dates of onset through the recollections of family members/caregivers and followed them forward to their dates of death. Thus, based on the prevalent sampling sheme, the observed failure times are deemed ``length-biased''. To distinguish biased and unbiased samples, we let $Y_1 \le \cdots \le Y_n$ denote the order statistics of a random sample of draws from a biased sampling distribution $G$, i.e., such that $G \neq F$.    

Analogous to Wilk's sample size problem given in Equation \eqref{sampsizeeq}, the problem for a generally biased/partially observed sample is based on solving for $n$ in the inequality
 \begin{equation}
\Pr\left[F(Y_{n-m+1}) - F(Y_r) \geq q\right] \geq 1 - \alpha.
\label{biassampsizeeq}
\end{equation}
To derive the required sample size $n$ based on a biased sample for a particular coverage on the target population distribution, we first consider the one-sided problems
\begin{equation}
\begin{aligned}
\Pr\left[F(Y_{r}) \le 1-q_r\right] & \geq 1 - \alpha_r, &
\Pr\left[F(Y_{n+1-m}) \ge q_m \right] & \geq 1 - \alpha_m,
\end{aligned}
\label{biasedcoveq}
\end{equation}
which correspond to placing a proportion $q_r$ ($q_m$) of the target population above (below) the $r$th smallest ($m$th largest) order statistic of the biased sample of size $n$ with probability $1-\alpha_r$ ($1-\alpha_m$).  Indeed, using a similar distributional argument as for obtaining~\eqref{betafree}, the CDF and inverse CDF of $F(Y_j)$ for $1 \le j \le n$ are given by
\begin{align}
H_j(z) := \Pr\left[F(Y_j) \le z\right] 
& = \Pr\left[G(Y_j) \le G(F^{-1}(z))\right] \nonumber \\
& = \pbeta(\Phi(z); j, n+1-j), \label{biasedcdf} \\
H_j^{-1}(p) & = \Phi^{-1}(\qbeta(p; j,n+1-j)), \label{biasedicdf}
\end{align}
where we define the quantile mappings $\Phi = G \circ F^{-1}$ and $\Phi^{-1} = F \circ G^{-1}$.  Equation~\eqref{biasedicdf} can be used to produce exact solutions to the one-sided coverage problems associated with~\eqref{biasedcoveq} in terms of the $\qbeta$ function.  Alternatively, one can replace the $\qbeta$ in these solutions by equating the righthand sides of~\eqref{unbiasedcov} and~\eqref{scheffecov}, thus giving the Scheff\'e-Tukey solutions to the one-sided coverage problems with biased sampling:
\begin{equation}
\begin{aligned}
    q_r & \cong 1 - \Phi^{-1} \left(1 - \frac{n - \frac{1}{2}(r-1) - \frac{1}{4} x_r}{n - \frac{1}{2}(r-1) + \frac{1}{4} x_r} \right), & 
    q_m & \cong \Phi^{-1} \left(1 - \frac{n - \frac{1}{2}(m-1) - \frac{1}{4} x_m}{n - \frac{1}{2}(m-1) + \frac{1}{4} x_m} \right),
    \end{aligned}
    \label{biasscheffecov1}
\end{equation}
where $x_r$ ($x_m$) is the $1-\alpha_r$ ($1-\alpha_m$) quantile of a Chi-Square distribution with $2r$ ($2m$) degrees of freedom.

\subsection{An Analytic Approach using Inequalities}\label{sec:ineq}

Consider the one-sided coverage problems associated with~\eqref{biasedcoveq} with coverage probabilities $1-\alpha_r = 1-\alpha_m = \sqrt{1-\alpha}$. Under the assumption that $Y_{r}$ and $Y_{n-m+1}$ are approximately independent for $n$ moderately larger than $r + m$,
it follows that  
\begin{equation}
\begin{aligned}
    \Pr\left[F(Y_{n-m+1}) - F(Y_r) \geq q_m - (1 - q_r) \right] 
    & \geq \Pr\left[ \left\{F(Y_r) \leq 1 - q_r\right\} \cap \left\{F(Y_{n-m+1}) \geq q_m\right\} \right] \\
    & \approx \Pr\left[F(Y_r) \leq 1 - q_r \right] \Pr\left[F(Y_{n-m+1}) \geq q_m \right] \\
    & \geq \left( \sqrt{1 - \alpha} \right)^2 = 1 - \alpha.
\end{aligned}
\end{equation}
Thus, using the Scheff\'e-Tukey formulas~\eqref{biasscheffecov1} for $q_r$ and $q_m$, a solution to the two-sided coverage problem with biased sampling for given $n$, $r$, $m$, and $\alpha$ is given by
\begin{equation}
q = q_m - (1- q_r) \cong \Phi^{-1} \left( \frac{n - \frac{1}{2}(m-1) - \frac{1}{4} x_m}{n - \frac{1}{2}(m-1) + \frac{1}{4} x_m} \right) - \Phi^{-1} \left( 1 - \frac{n - \frac{1}{2}(r - 1) - \frac{1}{4} x_r}{n - \frac{1}{2}(r - 1) + \frac{1}{4} x_r} \right).
\label{biasscheffecovfinal}
\end{equation}
Unlike in the unbiased setting of Section~\ref{sec:back}, the solution in Equation \eqref{biasscheffecovfinal} is not distribution-free and depends explicitly on $r$, $m$, and $\Phi^{-1} = F \circ G^{-1}$.  In most situations, the form of $\Phi^{-1}$ prevents isolating $n$ in Equation \eqref{biasscheffecovfinal} for an explicit sample size solution, as was possible with unbiased sampling. However, it is generally possible to approximate $\Phi^{-1}$ to very high accuracy by a piecewise linear function.  If this function has $M$ knots, then the righthand side of \eqref{biasscheffecovfinal} is a piecewise rational function of $n$ with at most $2M$ knots, for which we can find the interval $(n_L, n_U)$ containing the $n$ for solving \eqref{biasscheffecovfinal}, and then solve for $n$ on this interval by finding the roots of a quadratic.  Thus, a piecewise linear approximation of $\Phi^{-1}$ in~\eqref{biasscheffecovfinal} yields an analytic sample size formula at the cost of evaluating the right-hand size of \eqref{biasscheffecovfinal} at the $2M$ knots.

\subsection{A Computational Approach using the Fast Fourier Transform}\label{sec:fft}

In the inequality approach presented in Section~\ref{sec:ineq}, the determination of $q$ relied on a sequence of inequalities which provide loose bounds on the probability of the event $F(Y_{n-m+1}) - F(Y_r) \geq q_r + q_m - 1$, which leads to overestimation of the required sample size. Alternatively, once again assuming that $F(Y_{n-m+1})$ and $F(Y_r)$ are independent for $n$ moderately larger than $r + m$, an estimate of the probability distribution function (PDF) of $F(Y_{n-m+1}) - F(Y_r)$ is given by the convolution between their individual PDFs, which are readily obtained from the CDF formula in Equation~\ref{biasedcdf}.  In turn, this convolution can be efficiently calculated using the fast Fourier transform (FFT).

That is, let $H_r(z)$ and $H_{n+1-m}(z)$ denote the CDFs of $F(Y_r)$ and $F(Y_{n+1-m})$ given by~\eqref{biasedcdf}, and $h_r(z)$ and $h_{n+1-m}(z)$ denote the corresponding PDFs.  Then under the independence assumption above, the PDF of $F(Y_{n+1-m}) - F(Y_r)$ is given by
\begin{equation}
h_{r,m}^n(z) = \int_{-\infty}^{\infty} h_{n+1-m}(x) h_r(x-z) \, \mathrm{d}x.
    \label{convolu}
\end{equation}
We now seek to approximate this integral by a Riemann sum on a grid of size $\dx$.  Let $(L_r, U_r)$ denote an interval outside of which $h_r(z)$ is effectively zero.  These bounds can be obtained analytically using~\eqref{biasedicdf} via $L_r = H_r^{-1}(\epsilon/2)$ and $U_r = H_r^{-1}(1-\epsilon/2)$ for arbitrarily small $\epsilon$, such that $\Pr\left[F(Y_r) \notin (L_r, U_r)\right] = \epsilon$.  Let $(L_m, U_m)$ denote the corresponding interval for $h_{n+1-m}(z)$.  Without loss of generality, assume that the quantities

\begin{equation}
\begin{aligned}
k_0 & = (L_m - L_r)/\dx, & N_m & = (U_m - L_m)/\dx, & N_r & = (U_r - L_r)/\dx
\end{aligned}
\label{FFTintervals}
\end{equation}
are all integer-valued, and for $k \in \mathbb{Z}$, define the quantities
\begin{equation}
\begin{aligned}
a(k) & := \frac{H_{n+1-m}(L_m + (k + 1)\dx) - H_{n+1-m}(L_m + k\dx)}{\dx} \approx 
h_{n+1-m}(L_m + k \dx), \\
b(k) & := \frac{H_{r}(-L_m - ((k+1) - k_0)\dx) - H_{r}(-L_m - (k - k_0)\dx)}{\dx} \approx 
h_{r}(- L_m - (k-k_0) \dx).
\end{aligned}
\label{FFTdiscmass}
\end{equation}
Then we have $a(k) = 0$ and $b(k) = 0$ for $k < 0$ and for $k \ge M = N_m + N_r$.  Consequently, the Riemann sum approximation to $h_{r,m}^n(z)$ at $z = L_m + L_r + k \dx$ on the grid $\{L_m + j \dx: j \in \mathbb{Z}\}$ is given by
\begin{equation}
\begin{aligned}
h_{r,m}^n(L_m + L_r + k \dx) 
& \approx \sum_{j=-\infty}^\infty h_{n+1-m}(L_m + j \dx) h_r((L_m + j \dx) -(L_m + L_r + k \dx)) \dx \\
& = \sum_{j=-\infty}^\infty h_{n+1-m}(L_m + j \dx) h_r(-L_m - ((k-j) - k_0) \dx) \dx \\
& = \sum_{j=-\infty}^\infty a(j) b(k-j) \dx \\
% & = \sum_{j=0}^{M-1} a(m) b(k-m) \dx \\
& = \sum_{j=0}^{M-1} a(j) b((k-j) \mod M) \dx \\
& = \ifft(\fft(\bm{a}) \odot \fft(\bm{b})) \dx,
\end{aligned}
\label{fftconv}
\end{equation}
where $\bm{a} = (a(0), \ldots, a(M-1))$, $\bm{b} = (b(0), \ldots, b(M-1))$, the binary operator $\odot$ denotes the elementwise product between vectors, and $\fft()$ and $\ifft()$ are the discrete Fourier transform and its inverse, which can efficiently be computed in $\mathcal{O}(M \log M)$ using the FFT.  

Equation~\eqref{fftconv} is similar to the standard FFT-based convolution algorithm\cite{sakamoto.etal97,ruckdeschel.kohl14}.  However, the standard algorithm requires $h_r(z)$ and $h_{n+1-m}(z)$ to be evaluated on a common grid between $\min(L_m, L_r)$ and $\max(U_m, U_r)$, and for zero-padding to convert the linear convolution into a circular convolution to be performed via the FFT.  Thus, for the same grid size of $\dx$, the standard algorithm leads to FFTs of size at least $2\max\{N_1, N_2\} \ge M$, and of size considerably larger than $M$ when the supports of the $h_r(z)$ and $h_{n+1-m}(z)$ do not overlap.  

Given the grid-based approximation $\hat h_{r,m}^n(z)$ of $h_{r,m}^n(z)$, one can easily estimate its CDF $H_{r,m}^n(z)$ and inverse CDF $(H_{r,m}^n)^{-1}(p)$ via piecewise linear functions $\hat H_{r,m}^n(z)$ and $(\hat H_{r,m}^n)^{-1}(p)$.  The latter of these immediately produces a solution to the biased two-sided coverage problem defined by~\eqref{biassampsizeeq} via
\begin{equation}
    q \cong (\hat H_{r,m}^n)^{-1}(\alpha).
    \label{biasedcovtwo}
\end{equation}
For the sample size problem, the FFT-based approximation of $h_{r,m}^n(z)$ does not produce a closed-form solution.  However, it may be obtained via standard root-finding algorithms to solve for $n$ in~\eqref{biasedcovtwo}.  This approach is more computationally intensive than the inequality approach due to applications of the FFT at each step of the root-finding algorithm.  However, our simulation studies in Section~\ref{sec:sim} indicate that a solution for $n$ can typically be found in less than 0.1 seconds.

\subsection{Calculation of $\Phi$ and $\Phi^{-1}$} \label{sec:qmap}

Both the inequality approach and the computational approach depend on availability of the quantile mappings $\Phi = G \circ F^{-1}$ and $\Phi^{-1} = F \circ G^{-1}$.  In the context of sample size determination, this requirement is satisfied in a broad range of settings:
\begin{enumerate}
    \item Most sample size formulas depend explicitly on the exact specification of the target distribution\cite{lachin, liu, mcvit, patil2}, such that $F$ and $F^{-1}$ would typically be available by design.  In certain cases, such as the left-truncation setting we study in Section~\ref{sec:sim}, these produce analytic forms for $G$ and $G^{-1}$ as well.
    \item In more complex settings, both $F$ and $G$ (and their inverses) may be intractable, but sampling from $F$ and $G$ is straightforward.   In such cases, $F$ and $G$ (and their inverses) can be estimated via Monte Carlo.  These estimates are readily combined to form piecewise linear approximations of $\Phi$ and $\Phi^{-1}$ of arbitrary accuracy.  Moreover, this incurs only a one-time cost prior to the sample size calculation for either approach, and for any values of $r$, $m$, $q$, and $\alpha$.  This approach is illustrated for left-truncated and right-censored failure time data in Section~\ref{sec:sim}.
    \item In many applied settings, sample sizes are derived after an initial pilot study.  In such cases, a random sample from $G$ can be used to obtain empirical estimates of $G$ and $G^{-1}$.  Likewise, if the form of the bias is known, $F$ and $F^{-1}$ can be approximated by a nonparametric maximum likelihood estimator (NPMLE).  Once again, the corresponding $\Phi$ and $\Phi^{-1}$ can then be approximated by piecewise linear functions at a one-time cost.  This approach is illustrated in Section~\ref{sec:appl}.
\end{enumerate}  

\section{Simulation Study}\label{sec:sim}

We performed an extensive simulation study to (i) illustrate how the Scheff\'e and Tukey formula yields incorrect coverages when applied to biased samples and (ii) assess the performance of the proposed sample size and coverage formulae using the inequality and computational approaches. We fixed the parameters $\alpha = 0.05$, $q = 0.80$ and varied the values of $r = (1, 3, 5, 10)$ and $m = (1, 2, 7, 12)$. Thus, using these parameter values, we used the Scheff\'e-Tukey~\eqref{scheffesamp}, Inequality (Section~\ref{sec:ineq}), and FFT (Section~\ref{sec:fft}) approaches to estimate the smallest sample size $n$ required to satisfy
% generated samples of size $n$ using the various approaches such that the equation
\begin{equation}
\Pr\left[ F(Y_{n-m+1}) - F(Y_{r}) \geq 0.80 \right] \ge 0.95.
\label{simsetupeq}
\end{equation}
% was satisfied. 
In our simulations, the target population distribution $F$ was a Gamma distribution with parameters equal to $(2, 2)$, $(1, 2)$, and $(0.5, 2)$, in order to assess the sample size formulae for increasing, constant, and decreasing hazard rates.  The sampling distribution $G$ was specified through a combination of length-biased sampling and right-censoring of the forward failure times (times from the cross-section to failure) by independent Exponential times to allow for $0\%$, $10\%$, $25\%$ and $40\%$ censoring rates.  Thus, the length-biased failure times are informatively right-censored\cite{asgha}.

If $f(x)$ is the target population density, then the length-biased sampling density is $g(y) \propto y f(y)$.  Thus, in the absence of right-censoring, the quantile mappings $\Phi$ and $\Phi^{-1}$ for Gamma-distributed target populations can be computed analytically (Appendix~\ref{app:biasedgamma}).  In the presence of right-censoring, the quantile mappings are estimated via simulation as described in Section~\ref{sec:qmap}.

Performance of the sample size formulae is evaluated by drawing 1000 independent biased samples of the calculated size $n$, and computing the empirical probability $1-\hat \alpha$ that $F(Y_{n+1-m}) - F(Y_r)$, the proportion of the target population covered by the tolerance limits, exceeds the nominal coverage level:
\begin{equation}
1 - \hat{\alpha} = \frac{1}{1000} \sum_{j=1}^{1000} \mathbb{I}\left[ F(Y_{n-m+1,j}) - F(Y_{r,j}) \geq 0.80 \right].
\label{simprobest}
\end{equation}
The results of our experiments are presented in Figure~\ref{simfigure}.

Overall, the Scheff\'e-Tukey formula tended to underestimate the required sample size, with estimated coverage probability $1-\hat\alpha \ll 0.95$ in almost all cases.  
%incorrect coverage probabilities were determined by the Scheff\'e and Tukey formula in almost all cases. 
In contrast, the Inequality approach tended to overestimate the sample size, yielding coverage probabilities close to $1$ in all cases. The FFT approach outperformed both methods by correctly specifying the required sample size and obtaining empirical coverage probabilities close to $0.95$ in all cases. These results illustrate how the application of the Scheff\'e-Tukey formula to biased samples will not yield the correct sample sizes nor coverage values, whereas the Inequaliy approach tends to overestimate the required sample size needed to obtain the correct coverage. Furthermore, the Inequality approach tended to yield worse results as the proportion of right-censoring of the length-biased failure time data increased. While the FFT approach is based on approximating a density function through the fast fourier transform numerical procedure, the computational speed of these calculations is negligible and nearly as fast as the Scheff\'e-Tukey and Inequality approaches.

%We report the empirical coverage probabilities and average coverages in Table \ref{simtable1}. Overall, the incorrect coverage probabilities were determined by the Scheff\'e and Tukey formula  in almost all cases. In contrast, the inequality approach tended to overestimate the sample size yielding coverage probabilities close to $1$ in all cases. The computational approach outperformed both methods by correctly specifying the required sample size and obtaining empirical coverage probabilities close to $0.95$ in all cases. These results illustrate how the application of the Scheff\'e-Tukey formula to biased samples will not yield the correct sample sizes nor coverage values whereas the inequaliy approach tneds to overestimate the required sample size needed to obtain the correct coverage. Furthermore, the inequality approach tended to yield worse results as the proportion of right-censoring of the length-biased failure time data increased. While the computational approach is based on approximating a density function through the fast fourier transform numerical procedure, the computational speed of these calculations is negligible and nearly as fast as the Scheff\'e and Tukey and inequality approaches. For similar results in the constant hazard and decreasing hazard cases, refer to Tables \ref{simtable2} and \ref{simtable3} in Appendix~\ref{app:tables}.

\begin{figure*}[!htb]
\centerline{\includegraphics[scale=0.55]{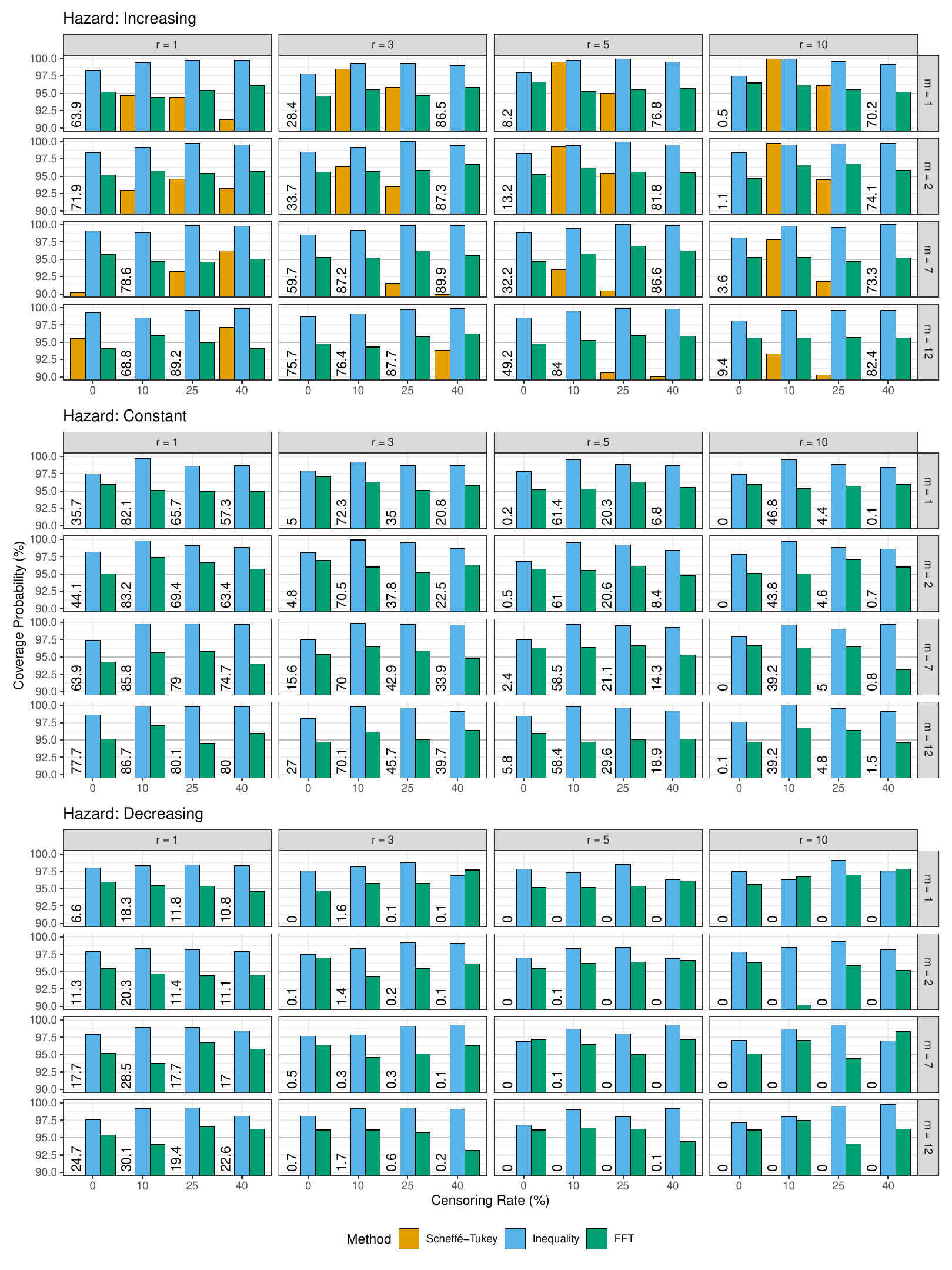}}
\caption{Empirical coverage probabilities ($1 - \hat{\alpha}$) computed using the Scheff\'e-Tukey formula, the Inequality approach or the FFT approach for length-biased right-censored data drawn from a Gamma distribution with either an increasing, constrant or decreasing hazard with varying censoring proportions over 1000 simulation replications. The values of $r$ vary over $(1, 3, 5, 10)$ and the values of $m$ vary over $(1, 2, 7, 12)$ with $q = 0.80$ and $1-\alpha = 0.95$.}
\label{simfigure}
\end{figure*}

\section{Application: Canadian Study of Health and Aging}\label{sec:appl}

The Canadian Study of Health and Aging (CSHA) was a multicentre study in Canada which initially recruited 14,026 subjects aged 65 years or older from the Canadian population\cite{wolfs}. For the approximately 10,263 subjects who agreed to participate, various exams were conducted to assess the subjects for vascular dementia or other dementias. A total of 822 subjects were positively screened for dementia and were classified into different dementia type groups for which approximately 21\% of the survival times were right-censored. 

Based on the prevalent cohort study design and the assumption that the CSHA subject onset dates are drawn according to a stationary Poisson point processs, the observed failure times were deemed length-biased. Thus, as alluded to in the simulation study (Section~\ref{sec:sim}), if the underlying failure time PDF is given by $f(x)$, then the length-biased PDF is given by $g(y) \propto y f(y) / \int_{0}^\infty t f(t)\, \mathrm{d}t$. Other examples of biased sampling schemes are outlined by Patil and Rao\cite{patil, patil2} and Patil et al.\cite{patil3}. Nonparametric maximum likelihood estimators (NPMLE) for the unbiased failure time distribution under biased sampling schemes have been thoroughly studied by Vardi\cite{vardi, vardi2, vardi3}, Gill et al.\cite{gill} and Vardi and Zhang\cite{vardi4}. When the length-biased failure time data are informatively right-censored, as in the case of the failure time data collected from the CSHA, Asgharian et al.\cite{asgha} and Asgharian and Wolfson\cite{asgha2} derived the unbiased failure time NPMLE and established its asymptotic properties. 

The empirical estimate $\hat G$ of the length-biased and right censored sampling distribution, and the NPMLE estimate $\hat F$ of the unbiased (target) failure time distribution are displayed in Figure~\ref{DAFig1}. 
% Consequently, we can estimate the length-biased distribution function $G(\cdot)$, the unbiased failure time distribution function $F(\cdot)$ and their respective inverse functions $\hat{G}^{-1}$, $\hat{F}^{-1}$. The empirical histograms are given in Figure 1.   
As indicated by the histograms in the Figure, the length-biased data tended to have a longer right-tail than the unbiased failure time data due to the selection of the longer failure times in the prevalent cohort study design.
\begin{figure*}[!htb]
\centerline{\includegraphics[scale=0.7]{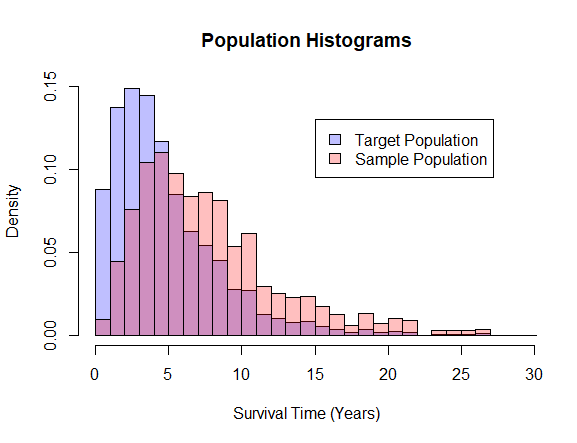}}
\caption{Empirical histograms of the target (unbiased - coloured blue) and sample (length-biased - coloured pink) CSHA populations based on the respective survival function NPMLEs.}
\label{DAFig1}
\end{figure*}
The CSHA data set is used as a pilot study to illustrate the third method of obtaining the quantile mappings described in Section~\ref{sec:qmap}.  With $\hat{G}$ and $\hat{F}$ the empirical and NMPLE estimates of the sampling and target distributions, respectively, we calculated the sample sizes for the Scheff\'e-Tukey, Inequality, and FFT approaches using the same values of $r = (1, 3, 5, 10)$, $m = (1, 2, 7, 12)$, $q = 0.80$, and $1-\alpha=0.95$ as in the simulation study of Section~\ref{sec:sim}.  These sample sizes are displayed in Table~\ref{DAtable}.  Granted, all of these sample sizes are smaller than the $n=821$ of the CSHA pilot study, meaning that none of the methods would have warranted any extra samples to be collected.  That being said, Table~\ref{DAtable} provides yet another perspective on the differences between the various approaches.  
\begin{table}[!htbp] \centering 
  \caption{Sample sizes computed using the Scheff\'e-Tukey, Inequality and FFT approaches based on using the CSHA data as a pilot study. The values of $r$ vary over $(1, 3, 5, 10)$ and the values of $m$ vary over $(1, 2, 7, 12)$ with $q = 0.80$ and $1-\alpha = 0.95$.} 
  \label{DAtable} 
\begin{tabular}{@{\extracolsep{5pt}} cc|c|c|c} 
\\[-1.8ex]\hline 
& & \multicolumn{3}{c}{Sample Size Methods} \\ \hline
r & m & Scheff\'e-Tukey & Inequality & FFT \\ \hline
1 & 1 & 22 & 98 & 69 \\
3 & 1 & 37 & 178 & 141 \\
5 & 1 & 50 & 247 & 204 \\
10 & 1 & 82 & 403 & 348 \\ \hline
1 & 2 & 30 & 104 & 72 \\
3 & 2 & 44 & 188 & 144 \\
5 & 2 & 57 & 263 & 207 \\
10 & 2 & 88 & 413 & 349 \\ \hline
1 & 7 & 63 & 130 & 91 \\
3 & 7 & 76 & 210 & 162 \\
5 & 7 & 88 & 289 & 225 \\
10 & 7 & 118 & 445 & 369 \\ \hline
1 & 12 & 94 & 158 & 111 \\
3 & 12 & 106 & 238 & 179 \\
5 & 12 & 118 & 308 & 242 \\
10 & 12 & 147 & 470 & 387 \\ \hline
\end{tabular}
\end{table}
We observe that the Scheff\'e-Tukey sample size formula yields relatively small values. We recall however that although these reported sample sizes are small, the Scheff\'e-Tukey sample size formula is independent of the data generating distribution and will therefore yield incorrect coverage values if a follow-up study to the CSHA were conducted. In contrast, the Inequality approach yielded much larger sample size values than both the Scheff\'e-Tukey and FFT approaches. The FFT approach produces sample sizes somewhere between the other two.  

To further illustrate this point, 
%was much larger than the Scheff\'e-Tukey as it incorporates the unbiased/biased data distributions however it did not provide sample sizes as large as the naive approach. 
we also compared the relative sample sizes of the three methods in the extreme order statistic setting $r = m = 1$, while varying the values of $q = (0.80, 0.85, 0.90, 0.95)$ and $1 - \alpha = (0.90, 0.925, 0.95)$. We present the results in Figure \ref{DAFig2}.  Once again Scheff\'e-Tukey and the Inequality approaches produce much smaller and larger sample sizes, respectively than the FFT approach, with the differences becoming larger with larger values of $q$.
\begin{figure*}[!htb]
\centerline{\includegraphics[scale=0.7]{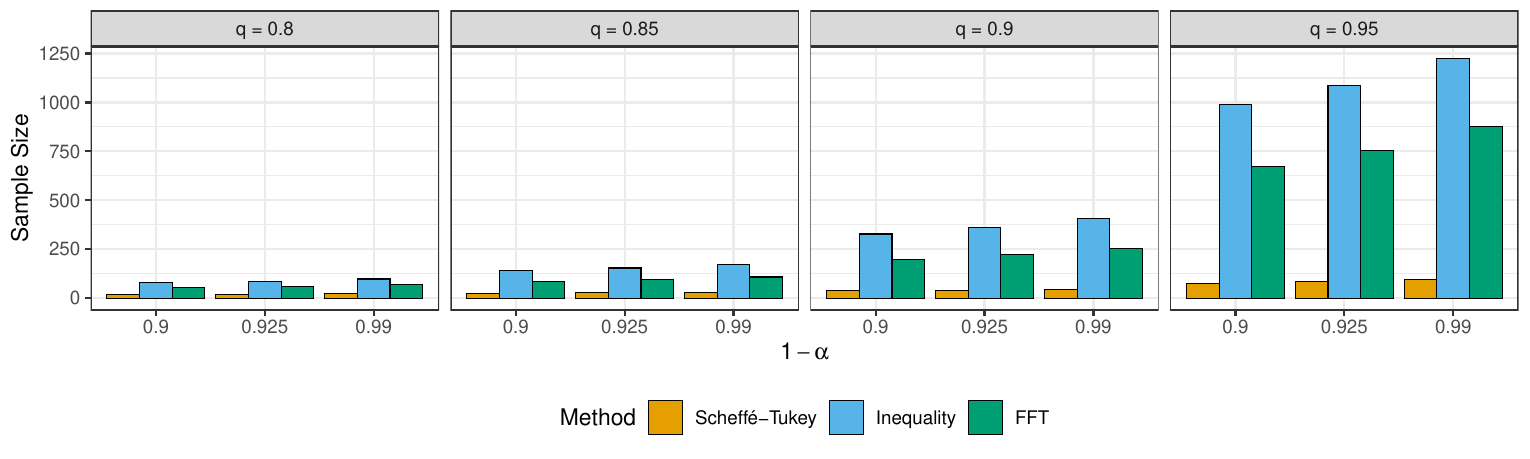}}
\caption{Sample sizes computed using the Scheff\'e-Tukey formula, the Inequality approach or the FFT Approach using the CSHA data as a pilot study. The values of $q$ vary over $(0.80, 0.85, 0.9, 0.95)$ and the values of $1-\alpha$ vary over $(0.9, 0.925, 0.95)$ with $r = m = 1$.}
\label{DAFig2}
\end{figure*}

\section{Discussion}\label{sec:disc}

The determination of a sufficient sample size to conduct a statistical hypothesis test at a preset level or to ensure the sample covers a preset portion of the population is crucial in the design of a given study. In this paper, we examined the effect of biased sampling schemes on the relation between sample size and population coverage through nonparametric tolerance limits determined by the order statistics of the data. For a large class of biased sampling schemes, including weighted distributions and censored samples, we showed how to estimate the required sample size using biased and partially observed data. Through a simulation study, we highlighted how the classical Scheff\'e-Tukey formula yields incorrect sample sizes and associated coverages and proposed two alternative approaches for biased and right-censored samples. These methods were based, respectively, on a simple probabilistic inequality and an FFT-based method for sampling directly from a difference in distributions of given order statistics. While the Inequality approach yielded overly large sample sizes with empirical coverage probabilities close to $1$, the FFT approach yielded the correct probabilities in all cases. Using the three separate sample size methodologies, we considered the length-biased right-censored survival times of participants in the Canadian Study of Health and Aging with a form of dementia as a pilot study to determine the required sample sizes for obtain appropriate coverage proportions between different order statistics. We obtained the same conclusions when comparing the three approaches as in the simulation study. Our practical recommendation is to employ the FFT approach whenever possible for computing tolerance limit sample sized with biased sampling.  Our experiments suggest that its computational cost relative to the Scheff\'e-Tukey and Inequality methods is a negligible tradeoff for the increase in accuracy for sample size determination in biased sampling regimes.

% In practice, the applied researcher can choose to use the inequality approach to obtain an upper bound on the required sample size to achieve the necessary coverage but then refine this sample size estimate using the computational approach to satisfy the logistical constraints in the study design.

%\backmatter
\bmsection*{Author contributions}
All authors contributed equally to this work.

\bmsection*{Acknowledgments}
The CSHA was supported by the Seniors Independence Research Program, through the National Health
Research and Development Program (NHRDP) of Health Canada (project 6606-3954-MC[S]). The progression of dementia project within the CSHA was supported by Pfizer Canada through the Health Activity Program of the Medical Research Council of Canada and the Pharmaceutical Manufacturers Association of Canada; by the NHRDP (project 6603-1417-302[R]); by Bayer; and by the British Columbia Health Research Foundation (projects 38 [93-2] and 34 [96-1]).
\

We acknowledge the support of the Natural Sciences and Engineering Research Council of Canada (NSERC), [J.H.M. - RGPIN-2024-04763, M.L. - RGPIN-2020-04364, M.A. - RGPIN-2024-05640].

\bmsection*{Financial disclosure}

None reported.

\bmsection*{Conflict of interest}

The authors declare no potential conflict of interests.

\bibliography{tolerance-statmed}

%\bmsection*{Supporting information}
%
%Additional supporting information may be found in the
%online version of the article at the publisher’s website.
\newpage
\appendix

\bmsection{Closure of Size-Biased Generalized Gamma Distribution}\label{app:biasedgamma}

Suppose $X$ is Generalized Gamma distributed with shape parameters $\alpha$, $\delta$ and rate parameter $\beta$ having density function, for $x > 0$, given by:
$$f(x; \alpha, \beta, \delta) = \frac{\delta \beta^\alpha}{\Gamma(\alpha/\delta)}x^{\alpha - 1} e^{-\left(\beta x\right)^\delta}.
$$
The Generalized Gamma distribution is particularly useful for survival analysis, covering many other distributions such as the Weibull, Gamma, Exponential, Half-Normal, Nakagami, and Fréchet.  

Under the general model of size-biased sampling~\cite{scheaffer72}, the resulting sampling distribution is given by
%Consider a k-th biased Generalized Gamma density function given by:
$$
g(x; \alpha, \beta, \delta) = \frac{x^\kappa f(x; \alpha, \beta)}{\mathbb{E}(X^\kappa)}
$$
for $\kappa > 0$. Rearranging terms, the above can be simplified to
$$
g(x; \alpha, \beta, \delta) = \frac{\delta\beta^{\alpha + \kappa}}{\Gamma((\alpha + \kappa)/\delta)} x^{\alpha + \kappa - 1} e^{-(\beta x)^\delta},
$$
which corresponds to a Generalized Gamma density function with shape parameters $\alpha + \kappa$, $\delta$ and rate parameter $\beta$. Thus, the Generalized Gamma distribution is closed under size-biased sampling of any degree $\kappa > 0$\cite{corre}. For a general discussion on the relationships between size-biased distributions of arbitrary degree, see Patil and Ord\cite{patil4}.
\newpage

\end{document}